# Phonon screening of excitons in atomically thin semiconductors

Woncheol Lee,[1] Antonios M. Alvertis,[2,3,4] Zhenglu Li,[4,3,5] Steven G. Louie,[3,4] Marina R. Filip,[6] Jeffrey B. Neaton,[3,4,7] and Emmanouil Kioupakis[8]

[1]Department of Electrical Engineering and Computer Science, University of Michigan, Ann Arbor, Michigan 48109, USA
[2]KBR, Inc. NASA Ames Research Center, Moffett Field, California 94035, United States
[3]Materials Sciences Division, Lawrence Berkeley National Laboratory, Berkeley, California 94720, USA
[4]Department of Physics, University of California Berkeley, Berkeley, California 94720, USA
[5]Mork Family Department of Chemical Engineering and Materials Science, University of Southern California, Los Angeles, California 90089, USA
[6]Department of Physics, University of Oxford, Oxford OX1 3PJ, United Kingdom
[7]Kavli Energy NanoScience Institute at Berkeley, Berkeley, 94720 California, USA
[8]Department of Materials Science and Engineering, University of Michigan, Ann Arbor, Michigan 48109, USA

**Abstract**

Atomically thin semiconductors, encompassing both 2D materials and quantum wells, exhibit a pronounced enhancement of excitonic effects due to geometric confinement. Consequently, these materials have become foundational platforms for the exploration and utilization of excitons. Recent *ab initio* studies have demonstrated that phonons can substantially screen electron-hole interactions in bulk semiconductors and strongly modify the properties of excitons. While excitonic properties of atomically thin semiconductors have been the subject of extensive theoretical investigations, the role of phonon screening on excitons in atomically thin structures remains unexplored. In this work, we demonstrate via *ab initio GW*-Bethe-Salpeter equation calculations that phonon screening can have a significant impact on optical excitations in atomically thin semiconductors. We further show that the degree of phonon screening can be tuned by structural engineering. We focus on atomically thin GaN quantum wells embedded in AlN and identify specific phonons in the surrounding material, AlN, that dramatically alter the lowest-lying exciton in monolayer GaN via screening. Our studies provide new intuition beyond standard models into the interplay among structural properties, phonon characteristics, and exciton properties in atomically thin semiconductors, and have implications for future experiments.

Atomically thin two-dimensional (2D) semiconductors, composed of a few atomic layers, are known to exhibit enhanced electron-electron and electron-hole interactions due to their intrinsic geometric confinement [1-4], leading to large exciton binding energies and light-matter interactions compared to their three-dimensional (3D) counterparts [5] and giving rise to intriguing phenomena such as exciton coherence and condensation [6, 7]. These characteristics have found device applications ranging from light-emitting diodes (LEDs) and solar cells [8-10] to excitonic switches and transistors [11-12]. The accurate determination of the exciton binding energy of atomically thin semiconductors, a central quantity that dictates the stability of excitons, is indispensable for a thorough understanding of excitonic processes in 2D materials.

In this context, *ab initio* many-body perturbation theory calculations within the *GW* approximation and the Bethe-Salpeter equation (BSE) approach have been employed to determine the exciton binding energy in various material systems, such as inorganic crystals [13-15], organic molecules [16], 2D materials [1, 17, 18], and heterostructures [19-21]. Most calculations with the *ab initio GW*-BSE method are performed in a limit where the nuclei are clamped and only account for the screening of the electron-hole Coulomb attraction by other electrons [13, 22], which in the static screening approximation corresponds to the optical dielectric constant ($\varepsilon_\infty$). However, phonons can play an important role in the screening of excitons [23-28]. In polar materials, longitudinal optical (LO) phonons can generate macroscopic polarization fields, which lead to an increased low-frequency (~THz) contribution to the dielectric constant beyond that of electrons [29]. Relatedly, an ad hoc effective dielectric constant, intermediate between optical and static dielectric constants, has been used in the Wannier-Mott model to interpret recent experiments [30].

Recent work introduced an approach for incorporating phonon-screening effects into calculations of excitonic properties [23], with Ref. [24] extending this to a fully *ab initio* approach, including finite temperature effects. These studies revealed that phonon screening significantly influences exciton binding energies in bulk polar materials, while highlighting their significant temperature dependence and novel screening mechanisms, such as via multiple LO modes or via acoustic modes. However, the effect of phonon screening on atomically thin semiconductors remains unexplored. While the exciton binding energy is usually much greater than typical phonon energies in low-dimensional material systems, suggesting that phonon screening may be safely neglected [31, 32], this assumption has not been tested with first-principles calculations. Moreover, the presence of encapsulating or barrier layers, a common feature in optoelectronic and excitonic devices involving 2D materials, can introduce a complex interplay between phonons and excitons that cannot be accounted for with simple models of electron-phonon interactions.

In this Letter, we investigate the temperature-dependent phonon screening of excitons in atomically thin semiconductors using first-principles $GW$-BSE calculations. We focus on atomically thin GaN quantum wells embedded within AlN quantum barriers, which have recently gained much interest due to the presence of quantum-confined carriers, efficient deep-UV emission, and room-temperature-stable excitons in a system composed of conventional semiconductor materials [33-35]. The ability to control the thickness of both quantum wells and barriers provides the opportunity to quantify the intricate interplay between geometrical confinement, excitonic effects, and phonon screening of excitons in low-dimensional systems. We demonstrate that the extent of phonon screening exhibits significant variability depending on the degree of quantum confinement, offering a pathway to finely tune exciton properties. The presence of quantum barriers results in the emergence of a phonon screening mechanism distinct from those observed in 3D bulk materials. Importantly, we show that the phonon screening properties of the quantum barriers or substrate can be equally important in determining the nature of low-lying excitons as that of electronic screening.

We investigate the phonon screening of excitons in atomically thin GaN quantum wells of varying thicknesses using *ab initio* many-body perturbation theory with the $GW$-BSE framework. Following Ref. [24], the finite-temperature contribution of phonon screening to the direct electron-hole kernel is expressed as:

$$K_{S,S'}^{ph}(\Omega, T) = - \sum_{c v \mathbf{k}; c' v' \mathbf{k}'; \nu} A_{c v \mathbf{k}}^{S*} g_{cc';\nu}(\mathbf{k}', \mathbf{q}) g_{vv';\nu}^{*}(\mathbf{k}', \mathbf{q}) A_{c'v'\mathbf{k}'}^{S'}$$

$$\times \left[ \frac{N_B(\omega_{\mathbf{q}\nu}, T) + 1}{\Omega - \Delta_{c'\mathbf{k}';v\mathbf{k}} - \omega_{\mathbf{q}\nu} + i\eta} + \frac{N_B(\omega_{\mathbf{q}\nu}, T) + 1}{\Omega - \Delta_{c\mathbf{k};v'\mathbf{k}'} - \omega_{\mathbf{q}\nu} + i\eta} \right]$$

$$+ \left[ \frac{N_B(\omega_{\mathbf{q}\nu}, T)}{\Omega - \Delta_{c'\mathbf{k}';v\mathbf{k}} + \omega_{\mathbf{q}\nu} + i\eta} + \frac{N_B(\omega_{\mathbf{q}\nu}, T)}{\Omega - \Delta_{c\mathbf{k};v'\mathbf{k}'} + \omega_{\mathbf{q}\nu} + i\eta} \right] \quad , \quad (1)$$

where $c$ and $v$ represent band indices for electrons and holes, respectively, while $\mathbf{k}$ and $\mathbf{q}$ denote the indices for electron and phonon momentum, respectively. $A_{cv\mathbf{k}}^{S} = \langle cv\mathbf{k}|S\rangle$ are the coefficients of the exciton wave function in state $S$, written in the free electron and hole basis; $g_{nm;\nu}(\mathbf{k}, \mathbf{q}) = \langle m\mathbf{k} + \mathbf{q}|\Delta\nu|n\mathbf{k}\rangle$ represents the electron-phonon matrix elements, where $\Delta\nu$ is the first-order

change of the self-consistent potential; $N_B$ the Bose-Einstein distribution of phonon modes at temperature $T$; and $\Delta_{c\mathbf{k};v'\mathbf{k}'} = E_{c\mathbf{k}} - E_{v'\mathbf{k}'}$ is the difference between single-particle excitation energies, computed here within the *GW* approximation. The first term represents screening of excitons due to phonon emission, while the second term, only relevant at finite temperature, is due to phonon absorption. We also highlight the connection of Eq. (1) to previous theoretical and experimental works on excitonic lines, as discussed in Ref. [36-38]. Specifically, we note that the imaginary part of Eq. (1) provides the rate of exciton dissociation into a free electron-hole pair, $\tau_S^{-1}(T) = 2|\text{Im}[K_{SS}^{ph}(\Omega_S, T)]|$ [39-42]. The scattering towards a free electron-hole pair contributes to the broadening of the absorption line, along with other processes like exciton-exciton scattering [36-38] and Auger-Meitner recombination. Additionally, the real part of the exciton-phonon self-energy, $\Delta E_B(T) = \text{Re}[K_{SS}^{ph}(\Omega_S, T)]$, provides the temperature-dependent renormalization of the exciton (binding) energy.

We perform *GW*-BSE calculations using the BerkeleyGW package [43,13,44] to obtain electron-hole amplitude values, interpolating the electron-hole kernel onto a dense Γ-centered patch in the Brillouin zone, which encompasses the region where the exciton coefficients $A_{cv\mathbf{k}}^S$ assume non-negligible values and which is necessary to achieve convergence of the exciton binding energy [15]. The electron-phonon matrix element calculations are computed within density-functional perturbation theory (DFPT) [45] on a coarse grid, and interpolated onto the same dense Γ-centered patch as for the exciton calculations, by employing Wannier-Fourier interpolation [46] using the EPW code [47]. (We note that while a more consistent description can in principle be achieved by using the *GW* perturbation theory [48], we relegate such calculations to future work.) Gauge consistency between the exciton coefficients and the electron-phonon matrix elements is ensured through the method described in Ref. [49] and Ref. [24]. Phonon screening modifies the BSE kernel, and we obtain the change in the exciton binding energy at temperature $T$ within perturbation theory, as $\Delta E_B(T) = \text{Re}[K_{SS}^{ph}(\Omega_S, T)]$. Computational details are given in the Supplemental Material. We note here that our approach is grounded within many-body perturbation theory and considers electron-phonon interactions to the lowest order, making it primarily applicable in the weak electron-phonon coupling regime [50]. The description of strong electron-phonon coupling and polaron formation, as discussed in Ref. [51], fall beyond the scope of our present work. However, it has been discussed elsewhere that accounting for the formation of electron- and hole-polarons, as well as the interference between these, could further contribute towards reducing the exciton binding energy at finite temperatures [50, 23]. Additionally, multi-phonon processes, which are not included in our study, can become significant near room temperature for certain systems [52].

The quasiparticle gap and exciton binding energy of atomically thin GaN quantum wells (QWs) embedded in AlN is plotted as a function of well thickness in Fig. 1. Optical absorption spectra without the inclusion of phonon-screening effects are presented in Supplementary Material (Fig. S2). Our calculation predicts excitonic emission from atomically GaN in the deep UV range, showing good agreement with the experimentally measured photoluminescence spectra (detailed comparison included in the Supporting Information). As the thickness of GaN is reduced to the monolayer (ML) limit, the quasiparticle band gap increases significantly compared to its bulk value of 3.5 eV due to quantum confinement [33-35]. Our calculations indicate that the exciton binding energy using the standard clamped-ion *ab initio GW*-BSE formalism without the inclusion of phonon screening effects (hereafter termed "bare") also increases as the thickness is reduced, reaching its maximum value of (0.24 eV) for the monolayer. The bare exciton binding energy in atomically thin GaN QWs is much higher than the value computed in bulk GaN (0.065 eV) [15], which can be attributed to geometrical confinement effects, as predicted analytically for quasi-2D excitons [53, 54]. We also note that these values are obtained when neglecting phonon screening, leading to a reduced screening of the Coulomb interaction and thus higher exciton binding energy when compared with the experimental value (0.021–0.028 eV) [55-57]. Using the first-principles phonon-screening formalism and computational workflow [23, 24], we found the exciton binding energy of bulk GaN to be 46 meV at 300K, which aligns more closely with experimental data [24]. We now extend our approach to evaluate the effect of phonon screening on excitons and obtain the exciton binding energy of the GaN QWs at room temperature (Fig. 1b). Our computed value for the binding energy of 1 ML GaN when including phonon screening is 0.160 eV at room temperature, and this value decreases to 0.104 eV as the well thickness is increased to 4 ML. The incorporation of phonon screening in the BSE formalism results in a significantly decreased binding energy, however the calculated exciton binding energy remains sufficiently high, demonstrating the room-temperature stability of excitons in these atomically thin GaN structures.

Next, we investigate the specific mechanism through which phonon screening affects the exciton binding energy in these atomically thin semiconductors. In Fig. 2(a), we illustrate the temperature dependence of the relative reduction in the exciton binding energy due to phonon screening for varying well thickness (see Fig. S3 for absolute magnitude of the reduction). Our findings suggest a monotonic decrease in the influence of phonon screening as the well thickness is reduced. This observation aligns with standard expectations that excitons with larger binding energies are less susceptible to the effects of phonon screening.

To further understand the microscopic mechanism behind the phonon screening of low-lying excitons in thin GaN layers embedded in AlN, we analyze the individual phonon mode contributions (Fig. 2b-2d). In Fig. 2b, we examine the contribution of the polar longitudinal optical (LO) phonon mode of GaN to the screening of excitons. Our findings reveal that the LO phonon along the in-plane direction is the primary mode governing phonon screening. We observe that, as in the case of bulk GaN [24], the LO phonon exhibits a pronounced influence on the phonon screening of excitons in thin GaN layers, even with their larger exciton binding energies. This can be in part attributed to the higher LO phonon frequency of 110 meV in the quantum well structures (Figure S6), compared to 84 meV in bulk GaN [24], primarily resulting from aluminum (Al) atom contributions. The interaction between charge carriers and polar LO modes is characterized by the Fröhlich-type interaction [58], which diverges at the long-wavelength limit ($q \to 0$) [59] and is expected to dominate electron-phonon interactions in polar materials [60-62]. Since the exciton wavefunction is highly localized in reciprocal space perpendicular to the GaN QW normal, only the electron-phonon interactions with small phonon momenta ($q \to 0$) effectively contribute to the screening, which corresponds to the long-wavelength limit where the LO phonon dominates. The contribution of the LO phonon mode exhibits the same pattern as the total phonon screening, as shown in Fig. 2(a), with excitons exhibiting stronger binding being less susceptible to phonon screening. These outcomes are in good agreement with previous analysis of bulk materials [23, 24]. Furthermore, Fig. 2(b) illustrates that at lower temperatures, the screening exhibits a weak dependence on temperature, and is dominated by phonon emission. As temperature rises, the phonon occupation term, $N_B$ in Eq. (1), becomes more pronounced, leading to stronger temperature dependence, which is driven by both phonon absorption and emission.

Interestingly, we find that the longitudinal acoustic (LA) phonons also contribute to the screening of excitons, albeit to a lesser extent than the LO phonons, as has also been found for bulk GaN [24]. These LA modes are associated with piezoelectric-type electron-phonon interactions [63, 64], a byproduct of the polar symmetry of the structure, which are typically weaker than the Fröhlich interaction. An intriguing observation is that the degree of phonon screening due to LA modes increases as the GaN thickness layer is increased. This outcome can be attributed to the fact that in our supercell calculations, the in-plane lattice constant is fixed at the value of bulk AlN $(a_{AlN} = 3.112 Å)$ [65], mimicking the conditions of pseudomorphic growth in experiments [33, 34]. As a result, strain is induced in the GaN layers, and the effect of strain becomes more pronounced as the thickness of the GaN layer increases. This, in turn, enhances the effect of the piezoelectric interaction, reflected by the increased magnitude of the electron-phonon matrix elements for LA phonons (see Fig. S4 of the Supplementary Material). In contrast to the screening by LO phonons, which exhibit a moderate temperature dependence, the screening by piezoelectric LA phonons is strongly influenced by temperature. This distinction arises due to the LA phonons being positioned in the low-frequency range of the phonon dispersion (Fig. S6). Consequently, as the temperature increases, LA phonons are more readily occupied and play a significant role in exciton screening, despite their lower frequency compared to LO phonons.

Beyond screening from LO and LA phonons, there are contributions from other phonons that cannot be attributed to either the Fröhlich or the piezoelectric interaction, as depicted in Fig. 2(d). To uncover the origins of the additional phonon screening, we analyze the displacement patterns associated with these phonon modes. We find that these phonons exhibit displacements purely along the polar *c*-axis direction, and their oscillations are confined within the AlN quantum barrier region, as shown in Fig. 3(a) (also see Fig. S5 of the Supplementary Material). We classify these modes as half-space (HS) polar optical modes, which are predicted to exist in interfaces, quantum wells, and superlattices [66-69]. These HS modes are primarily associated with high-frequency LO phonons of the AlN quantum barrier along the z-axis (Fig. S6). However, compared to bulk AlN LO phonons, which propagate over the entire volume, these HS modes remain confined only within the AlN barriers even in the $q \to 0$ limit, showing the characteristics of a standing wave with an effective finite momentum along the z-axis $(q_z = n\pi/L, n = \text{integer})$. The emergence of the quantized effective phonon momenta stems from the modulation of periodicity along the z-axis, in close analogy with the quantization of energy levels observed in electrons confined within a potential well [69]. As these HS modes are localized within the AlN barrier, their contribution to the screening of excitons in the GaN well highlights the importance of the vibrational characteristics of the quantum barrier or the substrate materials on the screening of excitons confined within the optically-active 2D layer.

In general, the interaction between HS modes associated with the quantum barrier and the charge carriers in a QW has been regarded as insignificant in the past [70]. This is mainly because the interaction had been previously characterized by how much the electronic wavefunctions of charge carriers extend into the barrier region [66, 69]. For relatively thick QWs (with a thickness of more than a few nm) and for sufficiently high energy barriers ensuring effective carrier confinement, the effect of HS modes is negligible because the electron wavefunctions associated with quantum-well states remain confined within the QW region. However, this situation differs when considering atomically thin wells. When the thickness of the GaN QW is reduced to the atomic-scale limit, a significant portion of the wavefunction of the quantum-well state extends into the AlN barrier region, as illustrated in Fig. 3(b) and 3(c) (also see Fig. S7 of the Supplementary Material). This is primarily attributed to the evanescent nature of the carrier wavefunction that arises from the finite energetic height of the quantum barrier and from the extreme quantum confinement, which leads to a sharp increase in the energy levels within the QW [33, 34]. The reduced energy difference between the barrier and the quantum-confined state enhances the spread of carrier wavefunctions into the barrier region.

To gain further insights on the effect of these HS modes, we analyze the electron-phonon and hole-phonon matrix elements associated with the HS modes, for electrons at the conduction band minimum and holes at the valence band maximum, as shown in Fig. 3(d) for the case of the electrons (also see Fig. S4(c) of the Supplementary Material). Firstly, the matrix elements exhibit a sharp increase in the long-wavelength limit ($q \to 0$), similar to a Fröhlich-type interaction. However, due to the introduction of a non-zero effective momentum along the z-axis ($q_z \to q_z + n\pi/L$), the electron-phonon matrix elements of HS modes do not diverge in the long-wavelength limit. The absence of a divergence here is consistent with recent findings regarding the non-diverging Fröhlich interaction in 2D slab calculations [71]. As a result, the electron-phonon matrix elements of HS modes show a less prominent increase at long wavelengths when compared to those of the LO phonon along the in-plane direction. Secondly, we observe that electrons exhibit a more pronounced coupling to the HS modes compared to holes in the 2 - 4 ML GaN supercell structures. We attribute this trend to the greater spatial extent of the electron wavefunction within the barrier region compared to the hole wavefunction, which is a consequence of the substantially lighter effective mass of electrons [72]. However, in the 1 ML GaN limit, holes display a slightly stronger interaction with HS modes than electrons. We interpret the 1 ML case as the extreme 2D limit, where the majority of the supercell behaves as a half-space, as illustrated in Fig. S5(a), and thereby a large part of the carrier wavefunctions are affected by HS modes. Lastly, we find that the influence of the HS modes becomes more pronounced as the thickness of the QW is reduced, in agreement with the prior discussion on the extent of the carrier wavefunction into the barriers. In summary, for these and other atomically thin semiconductor QW structures, we can expect from our calculations that the enhanced interaction between charge carriers within the QW and the HS modes within the quantum barrier or the substrate contributes significantly to the phonon screening of excitons.

Finally, we note that the presence of HS modes is anticipated in various systems with confinement, not limited to infinitely repeating supercells but also including finite-size van der Waals stacks and isolated quantum wells. Moreover, the strong contribution by HS modes in the surrounding materials agrees with recent findings that emphasize the importance of substrates or quantum barriers on the properties of atomically thin semiconductors. For example, substrate screening results in significant bandgap renormalization and a substantial reduction of the exciton binding energy in these systems [73-76]. Other studies have also explored the influence of substrate screening on exciton transport [77]. Our calculations demonstrate an additional mechanism through which the surrounding environment can impact the properties of atomically thin semiconductors and underscore the important role of the substrate or quantum barrier in shaping their electronic, optical, and excitonic characteristics.

In summary, we performed a microscopic analysis of the phonon screening of excitons in atomically thin semiconductors in quantum well structures. Our investigation revealed the crucial role played by HS modes from neighboring materials, alongside an amplified contribution from piezoelectric LA modes. Moreover, we found that the impact of the LO phonon on exciton screening is more pronounced in atomically thin GaN compared to bulk GaN, owing to the higher frequency of the LO phonon in this system. Such distinctive phonon screening attributes clearly differentiate atomically thin semiconductors from conventional 3D bulk materials. These findings highlight the importance of the substrate or barrier materials, as they offer control over quantum confinement, strain, and, as now demonstrated, phonon screening of excitons. Our findings open novel pathways for manipulating the properties of excitons in 2D materials and provide valuable insights for the development and engineering of nano-scale devices based on atomically thin semiconductors.

**Acknowledgement**

This work is supported by the Computational Materials Sciences Program funded by the U.S. Department of Energy (DOE) Office of Science, Basic Energy Sciences (BES) under Award No. DE-SC0020129, which supported the investigations of electron-phonon coupling in atomically thin nitride heterostructures, and by the Theory of Materials FWP, which provided the formalism and methods for the phonon-screening of excitons and the Center for Computational Study of Excited-State Phenomena in Energy Materials (C2SEPEM) as part of the Computational Materials Sciences Program, which provided advanced codes for excited-state calculations based on the BerkeleyGW software, at the Lawrence Berkeley National Laboratory, funded by the U.S. Department of Energy, Office of Science, Basic Energy Sciences, Materials Sciences and Engineering Division, under Contract No. DE-AC02-05CH11231. The work is also partially supported by the National Science Foundation under Grant No. OAC-2103991 in the development of interoperable software enabling the EPW and BerkeleyGW calculations with consistent gauge. Computational resources were provided by the National Energy Research Scientific Computing (NERSC) Center, a U.S. Department of Energy (DOE) Office of Science User Facility supported under contract No. DE-AC02-05CH11231. M.R.F. acknowledges support from the UK Engineering and Physical Sciences Research Council (EPSRC), grants EP/V010840/1; EP/X038777/1.

# Figures

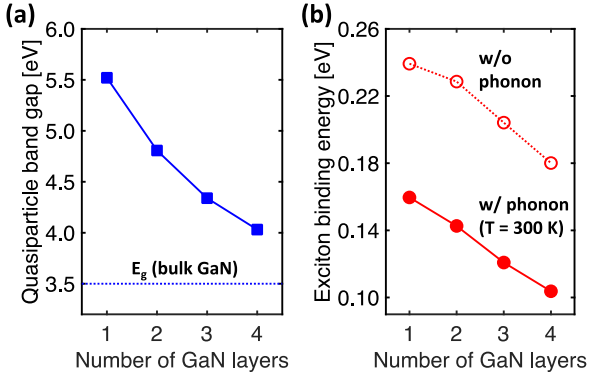

FIG. 1. (a) Calculated quasiparticle band gap of atomically thin GaN QWs in AlN barriers as a function of the number of GaN layers. (b) Calculated exciton binding energy of atomically thin GaN QWs in AlN barriers as a function of GaN well thickness, both with and without the inclusion of phonon-screening effects at 300 K within our BSE formalism.

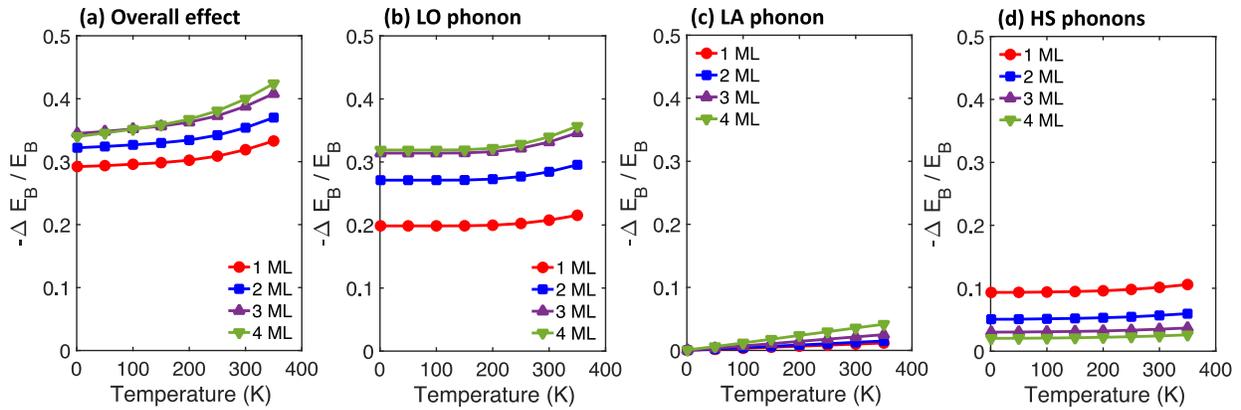

FIG. 2. (a) Relative reduction of the exciton binding energy as a function of temperature for atomically thin GaN QWs due to phonon screening. (b-d) The contribution to the reduction stemming from (a) polar LO phonons polarized along the in-plane direction, (c) in-plane LA phonons and (d) other phonon modes, categorized as HS modes.

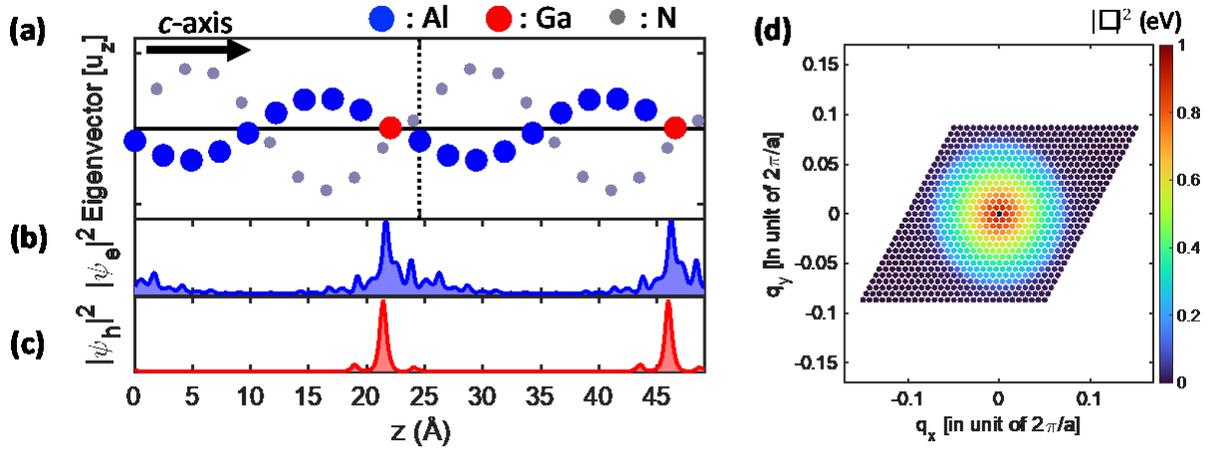

FIG. 3. (a) Eigenvector [$u_z$] of the highest-energy HS mode. Blue, red, and gray circles represent the displacement amplitude of Al, Ga, and N atoms along the z-axis, respectively. (b, c) Electron and hole wavefunctions averaged over the in-plane direction. (d) Magnitude of the electron-phonon matrix elements at the conduction band minimum for the highest-energy HS mode shown as a function of phonon momentum in the 2D Brillouin zone plane.

# Phonon screening of excitons in atomically thin semiconductors

## Supplemental Material

Woncheol Lee,[1] Antonios M. Alvertis,[2,3,4] Zhenglu Li,[4,3,5] Steven G. Louie,[3,4] Marina R. Filip,[6] Jeffrey B. Neaton,[3,4,7] and Emmanouil Kioupakis[8]

[1]Department of Electrical Engineering and Computer Science, University of Michigan, Ann Arbor, Michigan 48109, USA
[2]KBR, Inc. NASA Ames Research Center, Moffett Field, California 94035, United States
[3]Materials Sciences Division, Lawrence Berkeley National Laboratory, Berkeley, California 94720, USA
[4]Department of Physics, University of California Berkeley, Berkeley, California 94720, USA
[5]Mork Family Department of Chemical Engineering and Materials Science, University of Southern California, Los Angeles, California 90089, USA
[6]Department of Physics, University of Oxford, Oxford OX1 3PJ, United Kingdom
[7]Kavli Energy NanoScience Institute at Berkeley, Berkeley, 94720 California, USA
[8]Department of Materials Science and Engineering, University of Michigan, Ann Arbor, Michigan 48109, USA

## 1. Computational method

We performed first-principles calculations using density functional theory (DFT) and many-body perturbation theory (MBPT). We investigated a 20-atom wurtzite supercell structure consisting of atomically thin GaN quantum well with varying thickness (1 ML – 4 ML) embedded within AlN quantum barrier (FIG. S1). First, we performed structural relaxation calculation using local density approximation (LDA) for the exchange correlation functional [1-2], as implemented in Quantum ESPRESSO code [3]. Since the group-III nitrides require the $3d$ electrons for precise structural relaxations [4], we employed a norm-conserving pseudopotential of Ga, which include the $4s$, $4p$, and $3d$ electrons of Ga in the valence. Additionally, we fixed the in-plane lattice constant at the value of AlN ($a = 0.3112\text{Å}$) [5] and allowed atoms to move only along the polar $c$-axis of the supercell to mimic pseudo-morphic growth [6].

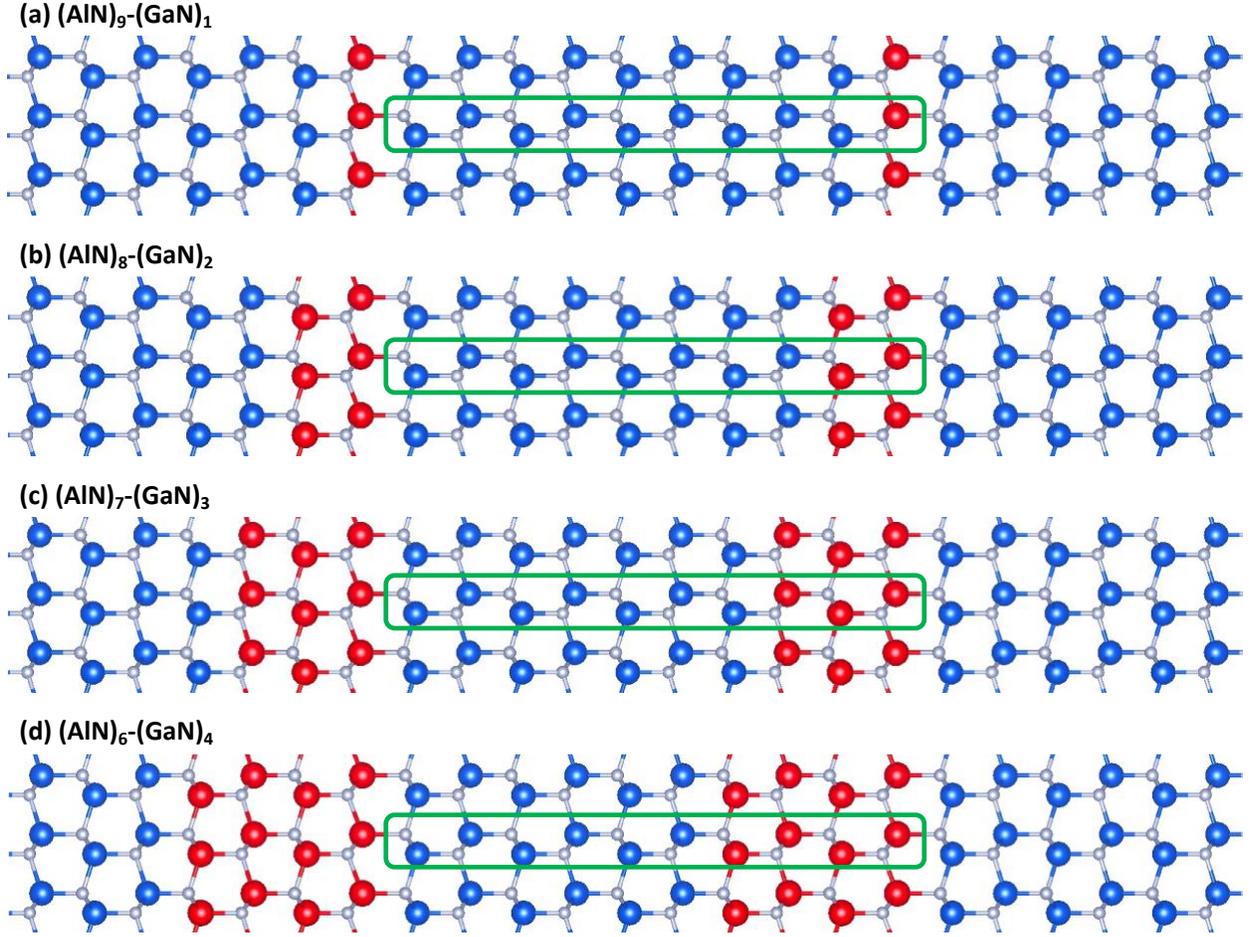

FIG. S1. Schematics of atomically thin GaN quantum wells with different thicknesses, embedded within AlN quantum barriers: (a) 1 ML GaN with 9 ML AlN, (b) 2 ML GaN with 8 ML GaN, (c) 3 ML GaN with 7 ML AlN, (d) 4 ML GaN with 6 ML AlN. The region highlighted by the green boundary represents the actual supercell employed in the DFT calculations. The circles in blue, red, and grey represent aluminum (Al), gallium (Ga), and nitrogen (N) atoms, respectively.

Next, we performed DFT calculations using Quantum Espresso [3]. We used a valence Ga pseudopotential that only consider $4s$ and $4p$ electrons to reduce the computation costs of the large-size supercell calculation. A planewave cutoff of 70 Ry with $8 \times 8 \times 1$ Monkhorst-Pack mesh converged the total energy of the supercell within 1mRy/atom.

Based on the DFT calculation, we calculated quasi-particle energy using $G_0W_0$ method, as implemented in the BerkeleyGW [7]. We used a screening cutoff energy of 20 Ry and $8 \times 8 \times 1$ Monkhorst-Pack mesh for the quasi-particle calculation. We chose plasmon-pole approximation to calculate the frequency-dependent dielectric response [8]. Also, we adopted the static-remainder approach to accelerate the convergence over the number of unoccupied electron states [9].

Finally, we used BerkeleyGW to solve Bethe-Salpeter Equation (BSE) [10]:

$$(E_{c\mathbf{k}+\mathbf{Q}}^{QP} - E_{v\mathbf{k}}^{QP})A_{vc\mathbf{k}}^S + \sum_{v'c'\mathbf{k}'} \langle vc\mathbf{k}|K^{eh}|v'c'\mathbf{k}'\rangle A_{v'c'\mathbf{k}'}^S = \Omega^S A_{vc\mathbf{k}}^S \qquad (1)$$

where $\mathbf{k}$ is an electron momentum, $\mathbf{Q}$ is an exciton center-of-mass momentum, $c$ and $v$ are conduction and valence band indexes, $E_{n\mathbf{k}}^{QP}$ is a quasi-particle energy, $K_{vc;v'c'}^{eh}$ is an electron-hole interaction kernel, $A_{vc\mathbf{k}}^S$ is an exciton wavefunction (i.e., electron-hole amplitude), and $\Omega^S$ is an exciton energy. In the calculations for 1 ML GaN and 2 ML GaN structures, we considered the lowest conduction band and the top three valence bands in the BSE calculation. In the case of 3 ML GaN and 4 ML GaN structures, we included the lowest conduction band and the top two valence bands. The same screening cutoff (20 Ry) was employed as in the quasiparticle calculation. For all cases, we use the patched sampling method [11] to interpolate the BSE kernel onto a Γ-centered patch of the Brillouin zone, drawn from a fine $140 \times 140 \times 1$ grid, converging the patch radius to ensure an accuracy of 10 meV or better for the exciton binding energy. After we obtain the exciton wavefunction ($A_{vc\mathbf{k}}^S$) from BSE calculations, we performed density-functional perturbation theory (DFPT) calculations to obtain the electron-phonon matrix [3]:

$$g_{ij,\nu}(\mathbf{k},\mathbf{q}) = \langle \psi_{i\mathbf{k}+\mathbf{q}}|\Delta_{\mathbf{q}\nu} v^{KS}|\psi_{j\mathbf{k}}\rangle, \qquad (2)$$

where $i$ and $j$ are band indexes, $\nu$ is a phonon branch index, $\mathbf{q}$ is a phonon momentum, $\psi_{i\mathbf{k}}$ is a Kohn-Sham wavefunction, and $\Delta_{\mathbf{q}\nu} v^{KS}$ is a phonon-induced variation of the Kohn-Sham potential. We employ Wannier-Fourier interpolation [12] to obtain the electron-phonon matrix on the same fine patch as for the exciton wavefunction. We use modified versions of the Wannier90 [13] and EPW [14] codes, in order to ensure the gauge consistency of the electron-phonon matrix computed from EPW and the exciton wavefunction obtained using BerkeleyGW. A detailed description of the computational workflow ensuring gauge consistency will be given elsewhere [15].

## 2. Absorption spectrum

Figure S2 shows the absorption spectra calculated for 1-4 MLs of GaN embedded within AlN barrier. Our calculation predicts a strong excitonic absorption peak appears in the deep UV range. We compare our calculated absorption spectra with experimentally measured photoluminescence spectra studied by several groups, thereby validating the reliability of our GW and BSE calculations.

For 1 ML (monolayer) GaN, which has been extensively studied due to its strong quantum confinement effect and deep-UV emission, multiple papers report a PL emission peak around 5.2-5.3 eV. This is consistent with our calculation results. Our simulated structure, with its in-plane lattice constant fixed to that of AlN to replicate actual pseudomorphic growth, shows the lowest 1s exciton peak at 5.28 eV (Figure S2a).

For 2 ML (bilayer) GaN, Refs. [16] and [17] reported a peak at 4.84-4.86 eV, which is slightly higher than our lowest 1s exciton peak at 4.58 eV (Figure S2, 2 ML GaN). This discrepancy may be due to thickness fluctuations in actual experiments, as discussed in Refs. [17] and [18], which is further evidenced by the relatively broad PL emission peak.

We denote that the absorption spectra calculations do not include phonon screening effects, which could renormalize the exciton binding energy as we discussed in the manuscript and cause slight shifts in the overall spectra.

| References | Photoluminescence emission peak |
| --- | --- |
| [16] Y. Taniyasu and M. Kasu Appl. Phys. Lett. **99**, 251112 (2011) | 0.9 ML GaN / 7.2 ML AlN: 5.23 eV<br>2 ML GaN / 8 ML AlN: 4.86 eV |
| [17] D. Bayerl *et. al.* Appl. Phys. Lett. **109**, 241102 (2016) | 1 ML GaN / 7 ML AlN: 5.32 eV<br>2 ML GaN / 8 ML AlN: 4.84 eV |
| [18] A. Aiello *et. al.* Nano Lett. **19**, 7852−7858 (2019) | 1 ML GaN / 10 nm AlN 5.18 – 5.28 eV |
| [19] A. A. Toropov et. al. Nano Lett. **20**, 158−165 (2020) | 1 – 2ML GaN: 5.17 – 5.39 eV |
| [20] Y. Wu *et. al.* Appl. Phys. Lett. **116**, 013101 (2020) | 1 ML GaN / 2 – 20 AlN: 4.9 – 5.25 eV |

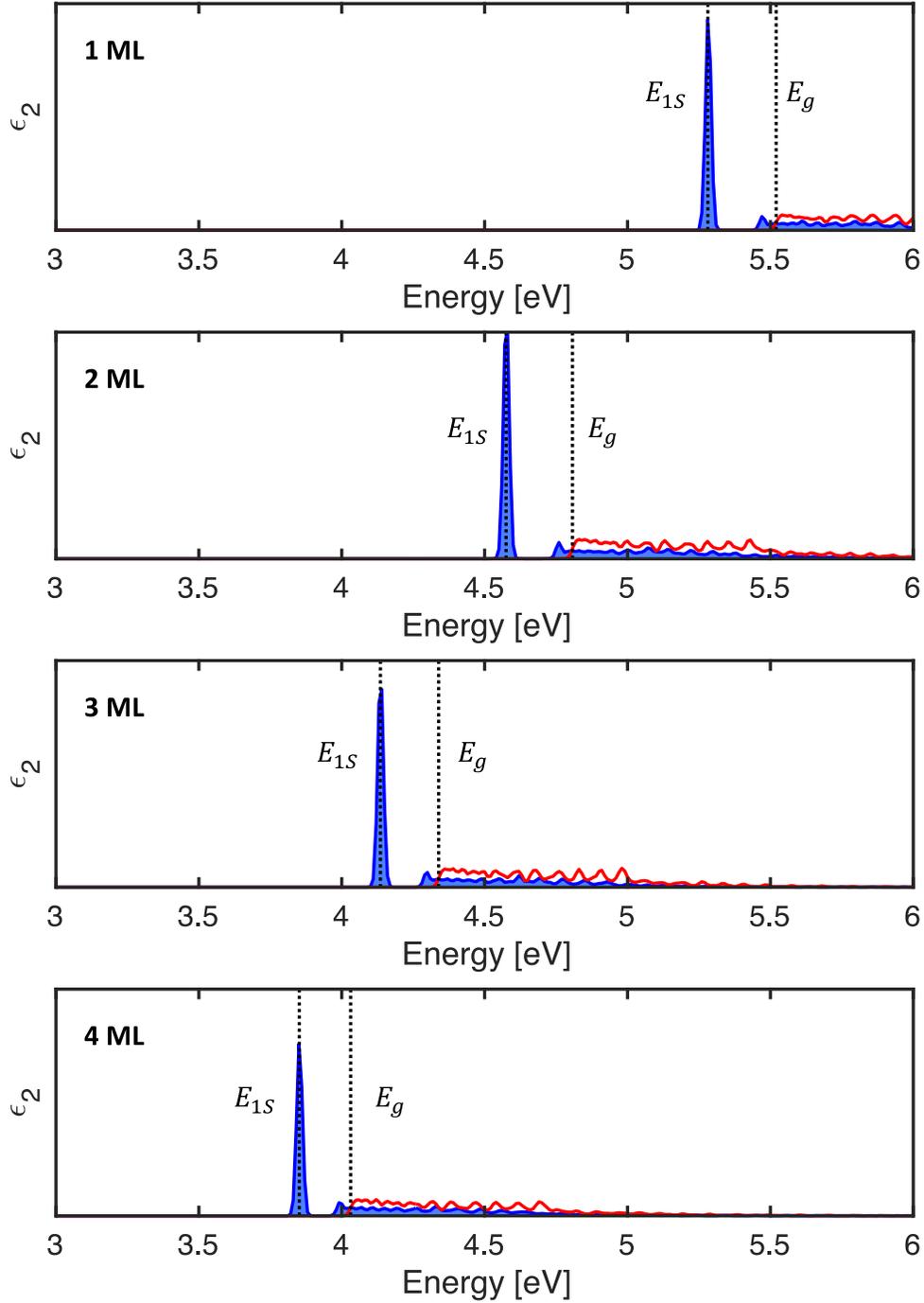

FIG. S2. Optical absorption spectra of atomically thin GaN with varying thicknesses. The blue shaded area indicates the spectra obtained from BSE, while the grey line represents the spectra obtained from random phase approximation (RPA). Vertical dashed lines indicate the energy of the lowest 1s exciton state ($E_{1s}$) from BSE calculations and the quasi-particle gap ($E_g$) obtained from $G_0W_0$ calculations.

### 3. Exciton binding energy obtained from the inclusion of phonon screening

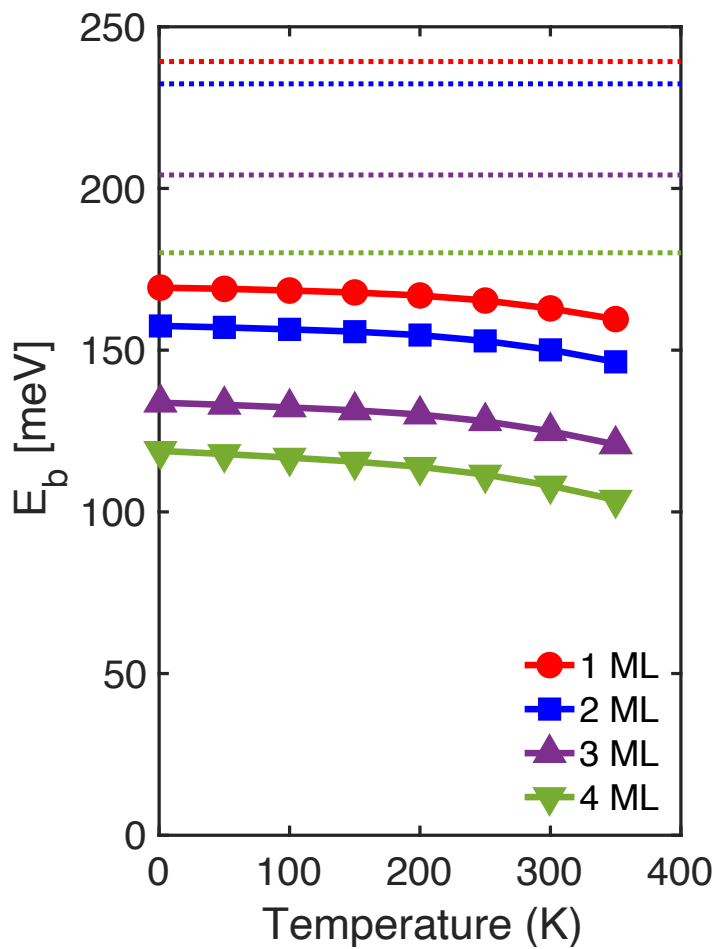

FIG. S3. Exciton binding energy calculated by the inclusion of the phonon screening effect (solid lines with symbols). Horizontal dashed lines indicate the exciton binding energy of four different GaN quantum wells, as determined from standard BSE calculation (without the inclusion of phonon screening).

## 4. Electron-phonon matrix

(a) LO phonon (Fröhlich interaction)

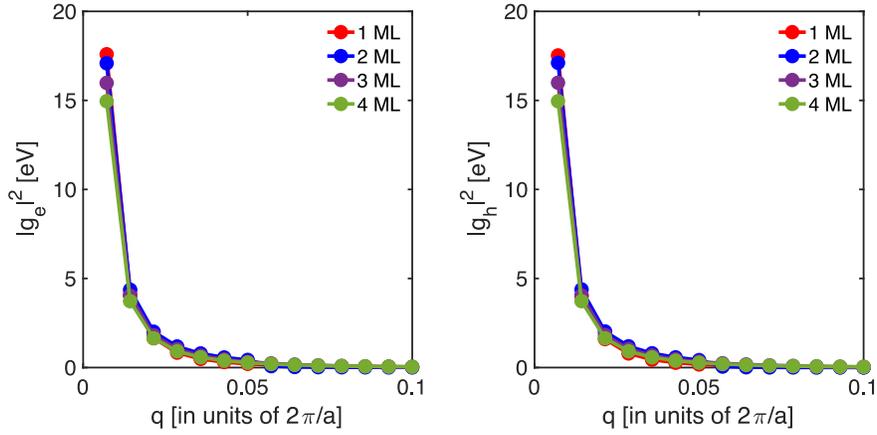

(b) LA phonon (Piezoelectric interaction)

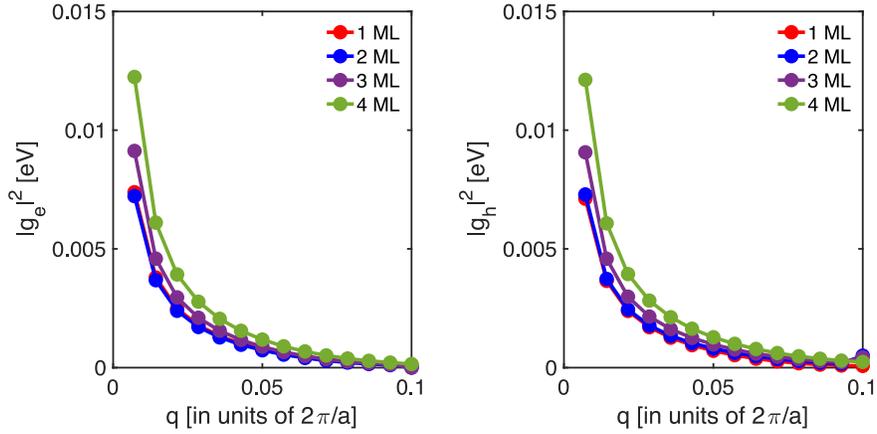

(c) HS phonons

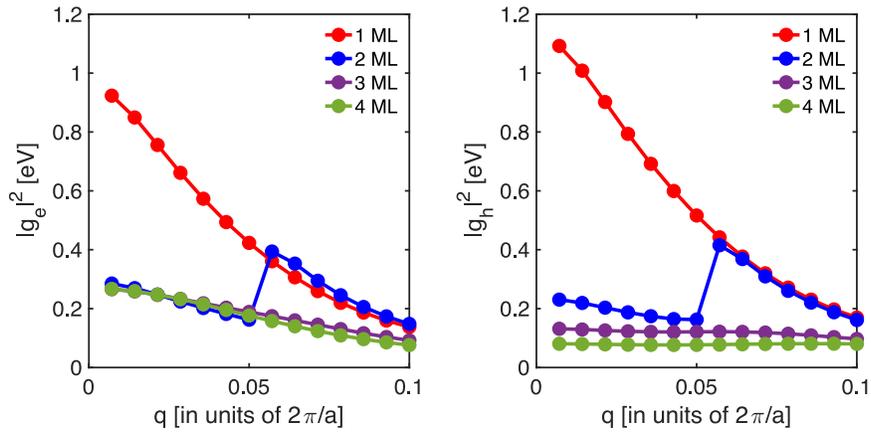

FIG. S4. Electron-phonon matrix $g_{cc',v}(\boldsymbol{k}=\Gamma,\boldsymbol{q})$ and hole-phonon matrix $g_{vv',v}(\boldsymbol{k}=\Gamma,\boldsymbol{q})$ for different vibrational modes. The discontinuity observed in the case of 2 ML arises from the crossing between LO and HS modes in the phonon dispersion.

## 5. Phonon eigenvector

**a) 1 ML GaN**

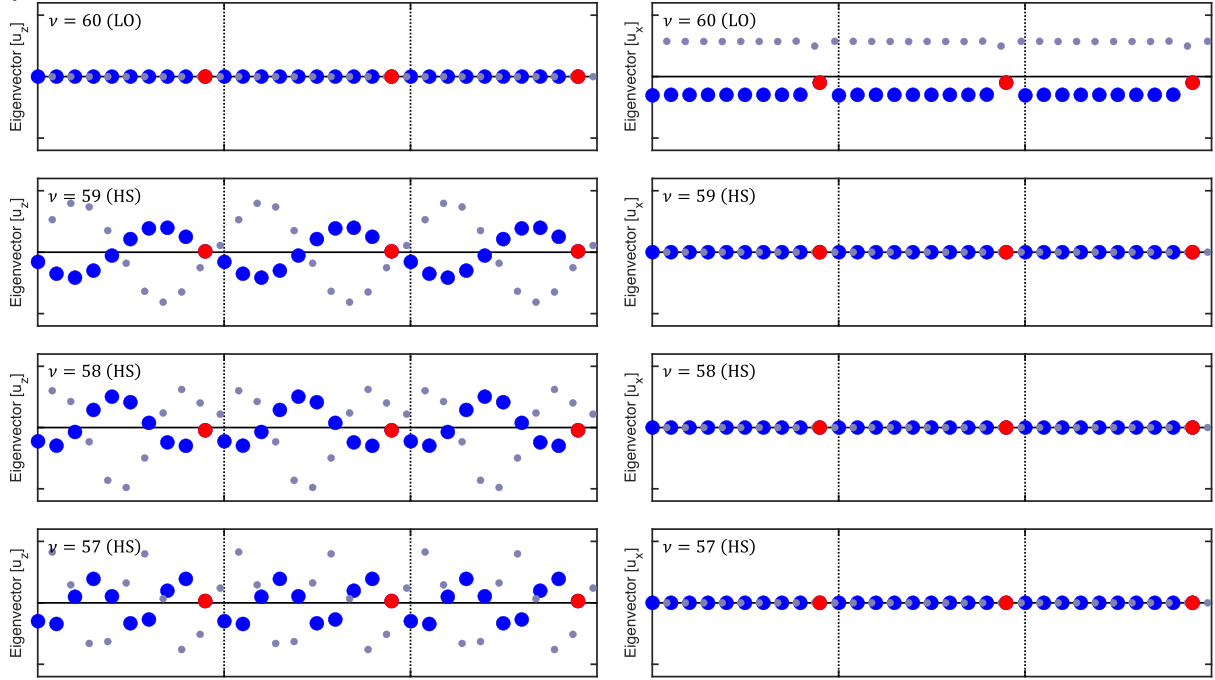

**b) 2 ML GaN**

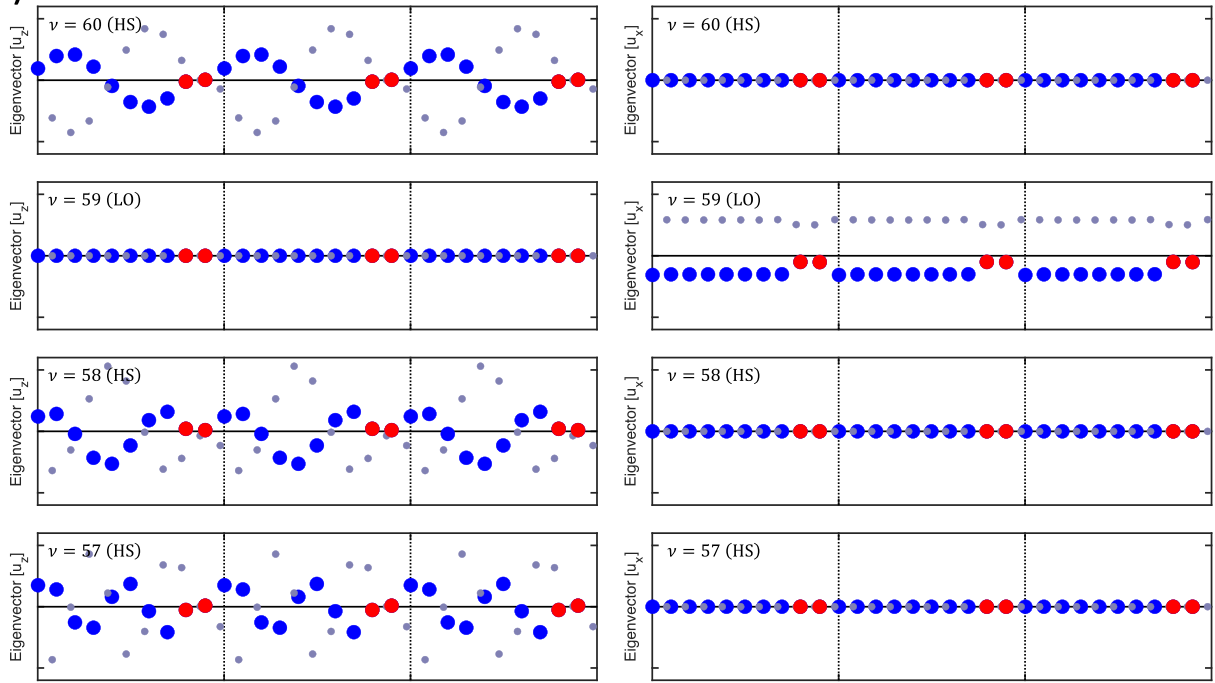

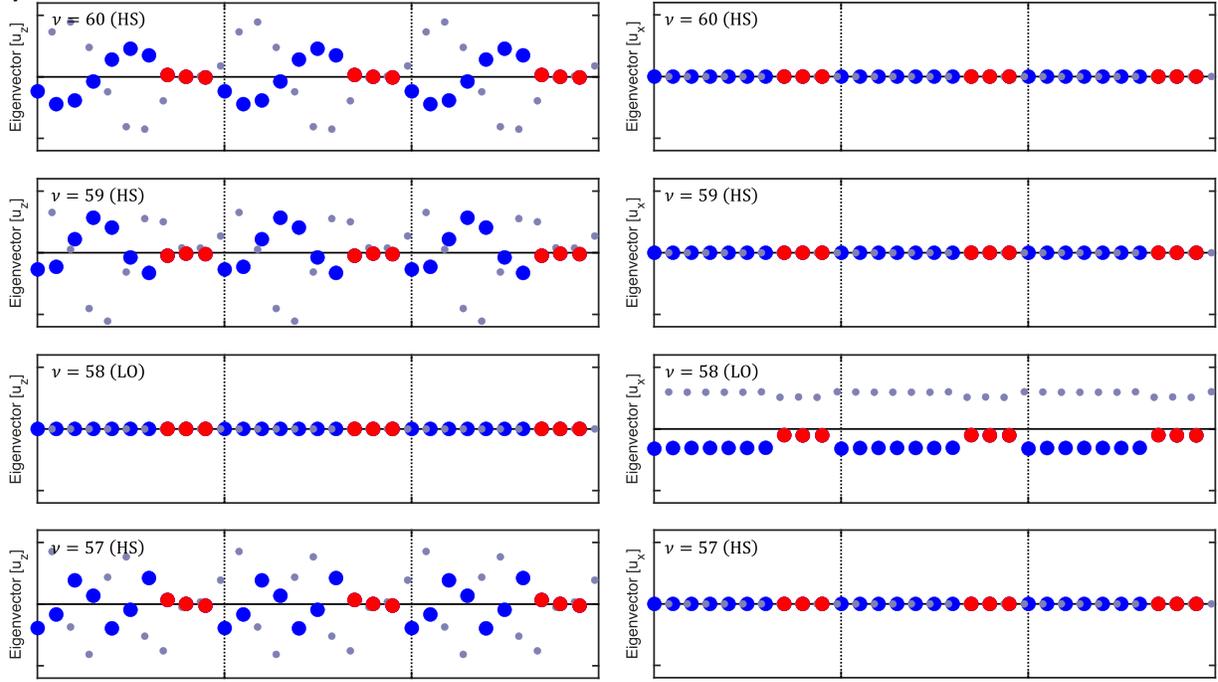
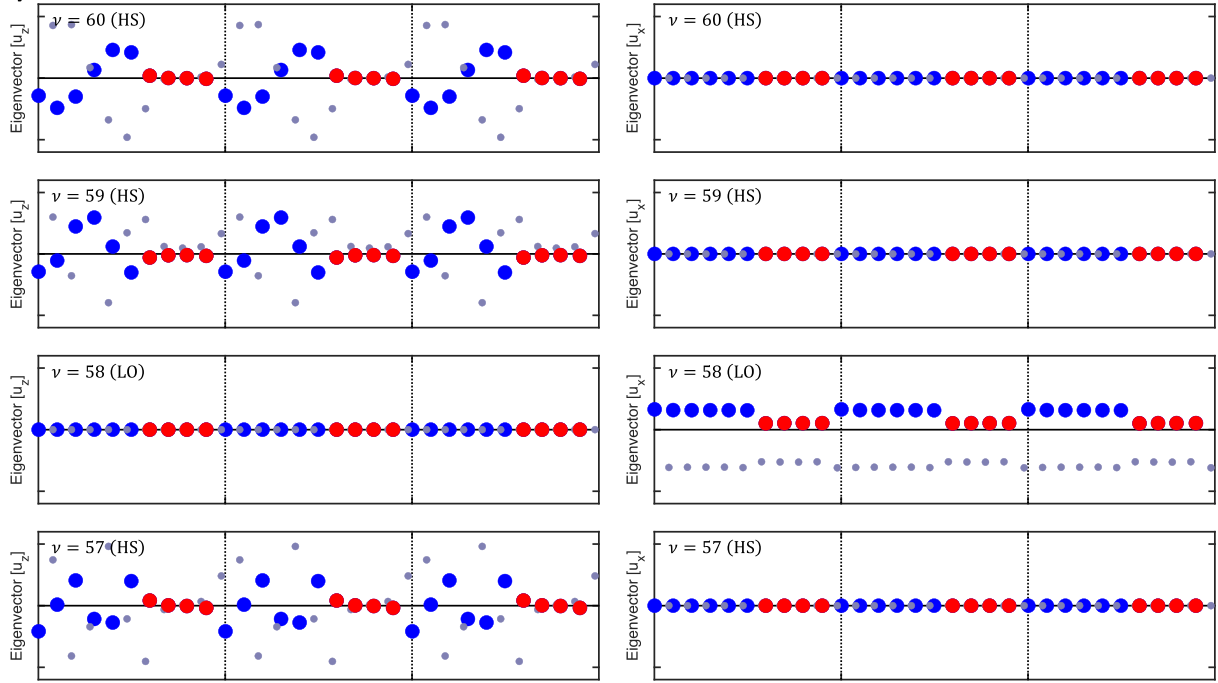

FIG. S5. Phonon eigenvectors shown for the LO phonon modes vibrating along the in-plane direction and the HS modes vibrating along the z-axis. For brevity, we are presenting only the four highest-energy phonon modes at $q = 0$.

## 6. Phonon dispersion

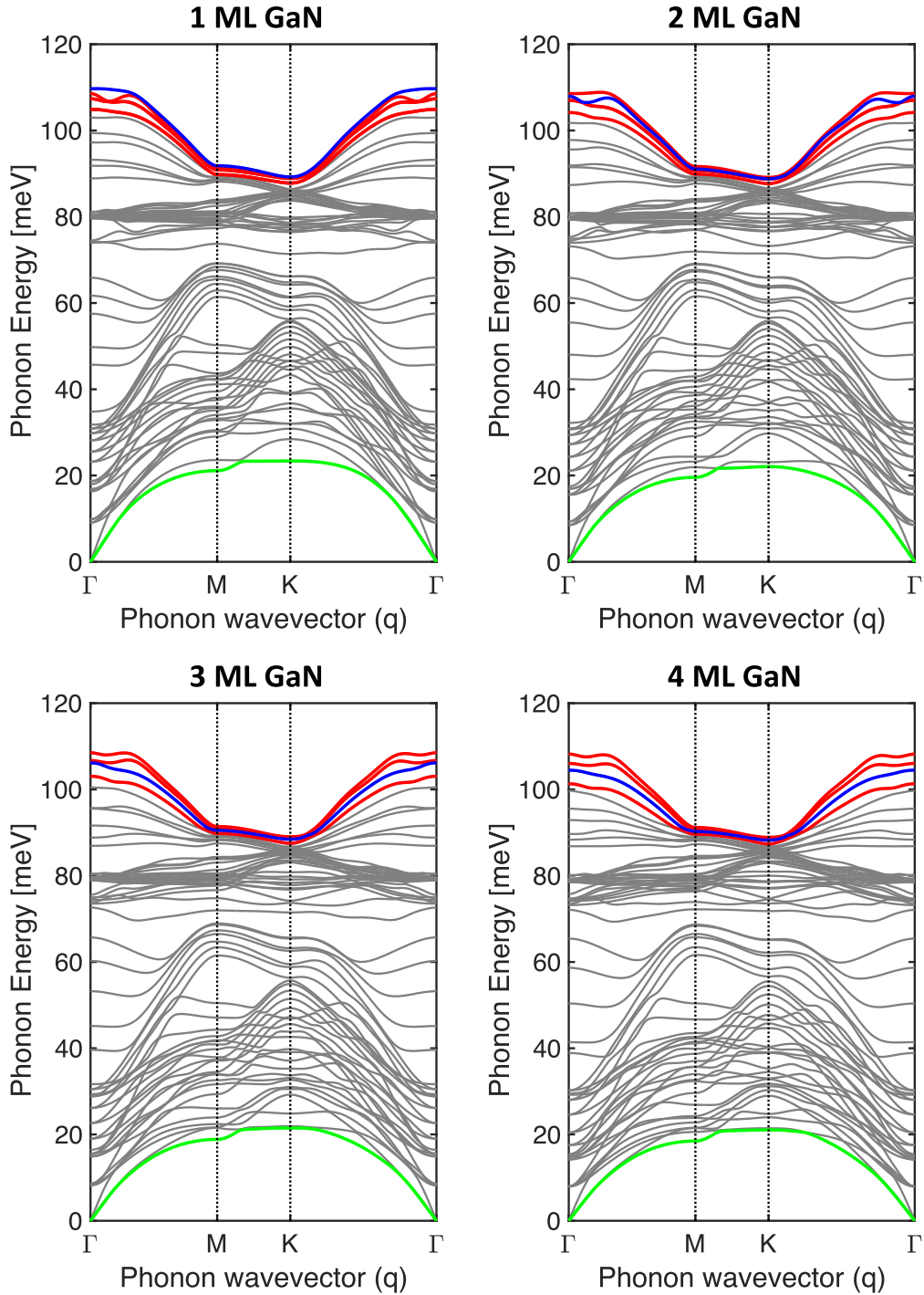

FIG. S6. Phonon dispersion of four different supercells, each with varying GaN quantum well thickness. The blue curve illustrates the Fröhlich-type LO phonon mode in the in-plane direction, the red curves indicate three HS phonon modes shown in Fig. S3, and the green curve represents the piezoelectric LA phonon mode.

## 7. Electron and hole wavefunctions

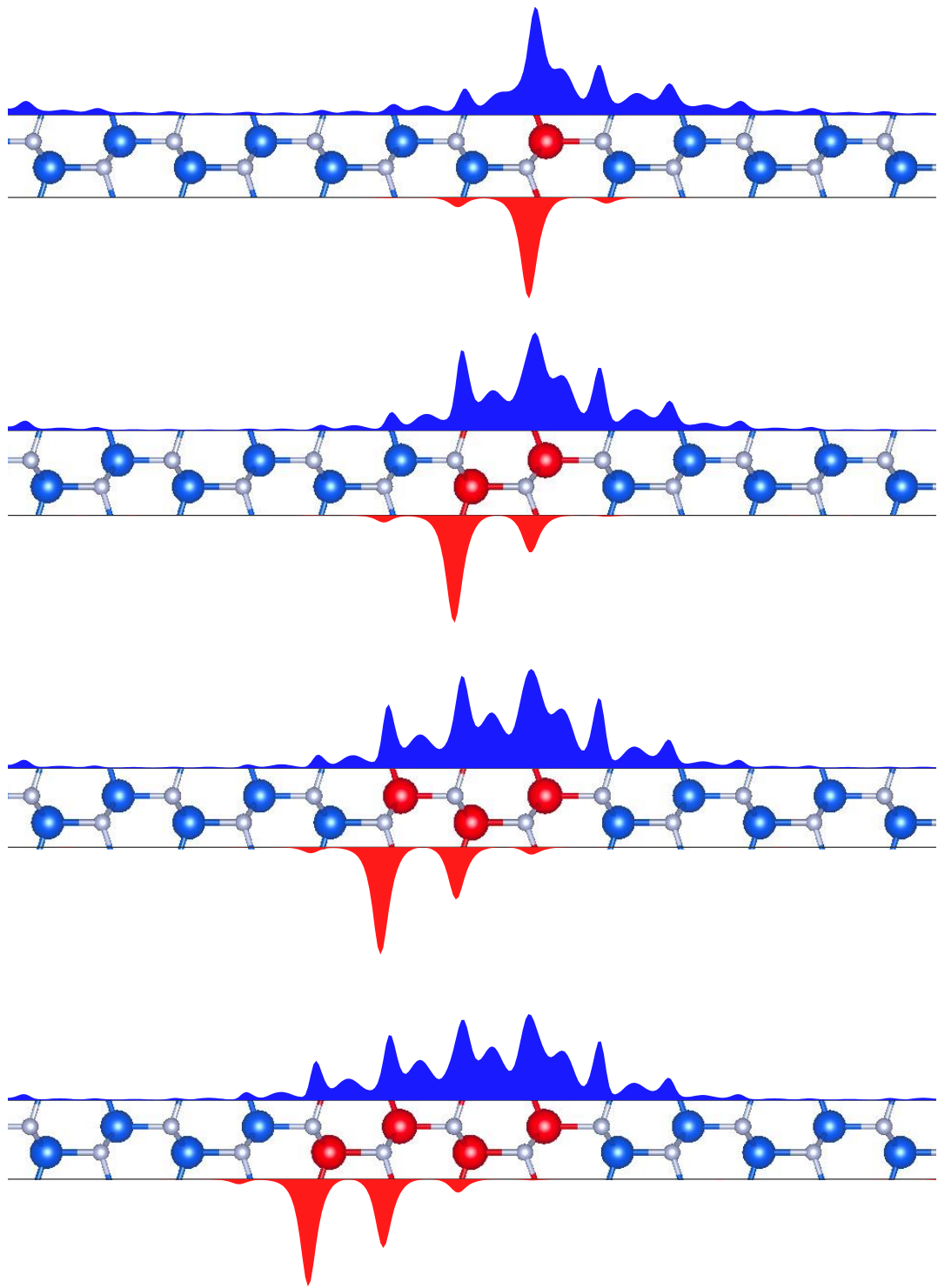

FIG. S7. Electron wavefunction (blue) and hole wavefunction (red) for 1-4 ML GaN quantum wells

## 8. Exciton wavefunction $A_{vck}$ in reciprocal space

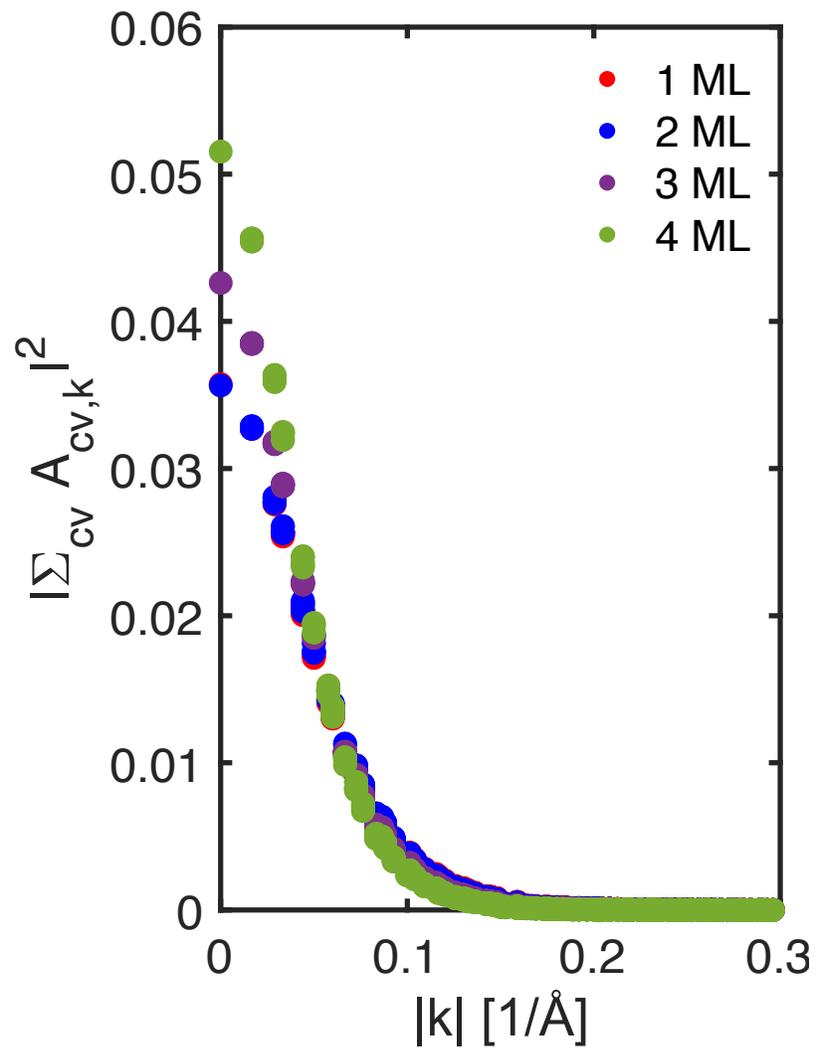